\documentclass[preprint]{emulateapj}
\usepackage{aas_macros}

\newcommand{\Caf}{SDSS J102915+172927}
\newcommand{\percc}{{\rm cm^{-3}}}
\newcommand{\um}{{\rm \mu m}}
\newcommand{\E}[1]{\times 10^{#1}}
\newcommand{\Pyroxene}{{\rm MgSiO_3}}
\newcommand{\Olivine}{{\rm Mg_2SiO_4}}
\newcommand{\Magnetite}{{\rm Fe_3O_4}}
\newcommand{\nH}{n_{{\rm H}}}

\newcommand{\Zsun}{Z_{\odot}}

\newcommand{\Dcrit}{{\cal D}_{{\rm crit}}}
\newcommand{\Zcrit}{Z_{{\rm crit}}}

\slugcomment{To Appear in ApJL}
\shortauthors{Chiaki, Nozawa, \& Yoshida}
\shorttitle{Grain growth in a low-metallicity gas}

\begin{document}

\title{Growth of dust grains in a low-metallicity gas \\
and its effect on the cloud fragmentation}

\author{
Gen Chiaki\altaffilmark{1},
Takaya Nozawa\altaffilmark{2},
and
Naoki Yoshida\altaffilmark{1,2}
} 
\altaffiltext{1}{
Department of Physics, Graduate School of Science, University of Tokyo, 
7-3-1 Hongo, Bunkyo, Tokyo 113-0033, Japan
}
\altaffiltext{2}{
Kavli Institute for the Physics and Mathematics of the Universe (WPI), 
Todai Institutes for Advanced Study, The University of Tokyo,
Kashiwa, Chiba 277-8583, Japan
}

\begin{abstract}
In a low-metallicity gas, rapid cooling by dust thermal emission 
is considered to induce cloud fragmentation and play a vital role in
the formation of low-mass stars ($\lesssim 1 \ M_{\odot }$) in 
metal-poor environments.
We investigate how the growth of dust grains through accretion of 
heavy elements in the gas phase onto grain surfaces alters the thermal 
evolution and fragmentation properties of a collapsing gas cloud.
We calculate directly grain growth and dust emission cooling 
in a self-consistent manner.
We show that $\Pyroxene $ grains grow sufficiently at 
gas densities $\nH = 10^{10}$, $10^{12}$, and $10^{14} \ \percc $ 
for metallicities $Z = 10^{-4}$, $10^{-5}$, and $10^{-6} \ \Zsun$, respectively,
where the cooling of the collapsing gas cloud is enhanced.
The condition for efficient dust cooling is
insensitive to the initial condensation factor of pre-existing grains
within the realistic range of 0.001--0.1,
but sensitive to metallicity.
The critical metallicity is $\Zcrit \sim 10^{-5.5} \ \Zsun$
for the initial grain radius $r_{\Pyroxene ,0} \lesssim 0.01 \ \um $ and
$\Zcrit \sim 10^{-4.5} \ \Zsun $ for $r_{\Pyroxene ,0} \gtrsim 0.1 \ \um $.
The formation of a recently discovered low-mass star with extremely low 
metallicity ($\leq 4.5 \E{-5} \ \Zsun$) 
could have been triggered by grain growth.
\end{abstract}



\section{INTRODUCTION}

Dust grains in the early universe are considered to be crucial for 
low-mass star formation in the low-metallicity gas.\footnote{
In this Letter, we use the term low-mass for a mass less than 1 $M_{\odot}$.}
The efficient cooling by dust thermal emission 
can trigger the fragmentation of the gas at densities 
$n_{{\rm H}} \sim 10^{13}$--$10^{15} \ \percc $ and
gas temperature $T_{{\rm g}}$ $\sim$ 1000 K, 
where the corresponding Jeans mass is less than one solar mass
\citep{Schneider03,Omukai05}.
In the early universe, dust grains are considered to be 
predominantly supplied by core-collapse supernovae; grains
are formed in the supernova ejecta \citep{Todini01,Nozawa03}
but a part of them are destroyed by the reverse shock 
penetrating into the ejecta \citep{Nozawa07,Bianchi07}.
Thus, the depletion factor of dust, $f_{{\rm dep}}$ (the mass ratio 
of dust to metals), is determined by the balance between 
the formation and destruction of grains.

\citet{Schneider12Crit} followed the thermal evolution of gas clouds 
adopting the depletion factors for their models of dust formation and destruction in supernovae.
For metallicity $Z$, the absolute amount of dust is related to 
the dust-to-gas mass ratio, ${\cal D} = f_{{\rm dep}} Z$.
\citet{Schneider12Crit} showed that the condition of cloud fragmentation
owing to rapid dust cooling is described as 
the minimal dust-to-gas mass ratio, 
${\cal D_{\rm crit}}$,
under the assumption that the dust-to-gas mass ratio 
is constant in the collapsing cloud.
However, \citet{Nozawa12} recently argued that the dust abundance can be 
increased by the growth of dust grains through 
accretion of heavy elements in the gas phase onto grain surfaces.
They found that the condensation efficiencies (the number fraction of an 
element depleted onto grains) of Si- and Fe-grains increase up to unity at gas densities
$n_{{\rm H}} =10^{10}$--$10^{14} \ \percc $ for
elemental abundances $[{\rm Si,Fe}/{\rm H}]=-5$ to $-3$.
However, they did not follow 
the thermal evolution of the gas clouds 
to determine whether the conditions for fragmentation are met in the clouds. 
In this Letter,
we investigate the effects of grain growth on the 
thermal evolution of star-forming clouds by calculating
grain growth, formation of hydrogen molecules on the grain surfaces,
and dust cooling self-consistently.
Then, by investigating possible range of metallicities and initial condensation
efficiency factors,
we can determine the critical metallicity, the minimal metallicity required for
the formation of low-mass fragments.

\begin{figure}[b]
\epsscale{1.0}
\plotone{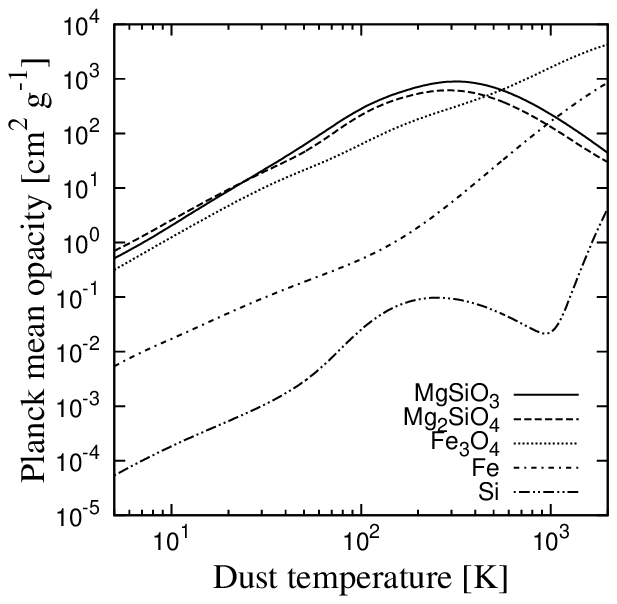}
\caption{
Planck mean opacities per unit dust mass for
$\Pyroxene $ (solid), $\Olivine $ (dashed),
$\Magnetite $ (dotted), Fe (dot-dashed), and Si (dot-dot-dashed)
as a function of dust temperature for 
a grain radius $0.01 \ \um$.
The values are insensitive to the grain radius except 
for Fe and $\Magnetite$.
}
\label{fig:Tkappa}
\end{figure}

\begin{figure}[t]
\epsscale{1.00}
\plotone{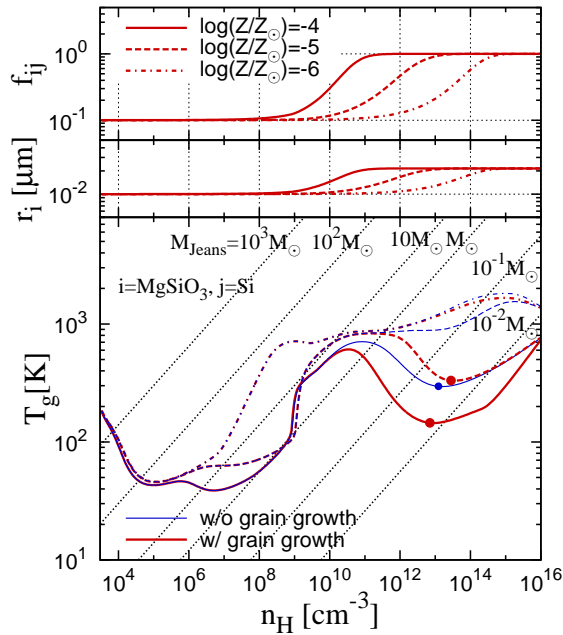}
\caption{
Condensation efficiency (top) and 
the radius of $\Pyroxene $ grains (middle), and 
gas temperature (bottom)
as a function of hydrogen number density
in gas clouds with metallicities 
$10^{-4} \ \Zsun$ (solid), 
$10^{-5} \ \Zsun$ (dotted), and 
$10^{-6} \ \Zsun$ (dot-dashed)
for the initial condensation factor 0.1 and
the initial grain radius 0.01 $\um$.
With (red) and without (blue) grain growth.
Filled circles mark the states where the 
fragmentation conditions are met (see the text).
}
\label{fig:nT}
\end{figure}

\begin{figure*}[t]
\epsscale{0.80}
\plotone{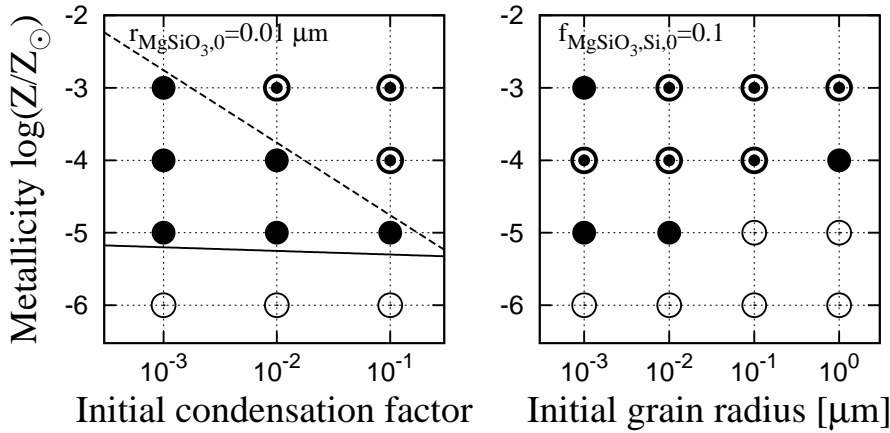}
\caption{
Fragmentation properties of the collapsing gas clouds for each model
on $Z$-$f_{\Pyroxene ,{\rm Si},0}$ (left) 
and $Z$-$r_{\Pyroxene ,0}$ (right) planes.
For models marked by double circles,
the fragmentation conditions are met both with and without grain growth.
While, for models marked by filled circles,
the conditions are met only with grain growth.
For models marked with open circles, 
the conditions are not met both with and without grain growth.
In left panel, the dashed line shows
the metallicity corresponding to the critical 
dust-to-gas mass ratio given by 
\citet{Schneider12Crit} (see the text) and the solid line shows 
the critical metallicities obtained by detailed calculations.
}
\label{fig:ox}
\end{figure*}

\section{NUMERICAL METHOD}
\label{sec:NumMethod}

We explore the thermal evolution of a collapsing gas cloud with 
metallicities of $Z = 10^{-6}$--$10^{-3}$ $Z_\odot$.
The evolutions of density $\rho$ and gas temperature 
$T_{{\rm g}}$
of the 
cloud core are followed by a one-zone model. Our calculations include 
chemical reactions of gas species and radiative transfer 
for line emissions from the gas, and continuum 
emissions from the gas and dust \citep[see][for details]{Chiaki12}. 

In our calculations where the growth of dust grains is included,
the mass density of dust $\rho _{{\rm d}}$ increases with time.
Thus, we calculate the continuum optical depth 
as follows;
\begin{equation}
\tau _{{\rm cont}}= \left( \kappa _{{\rm g}} \rho _{{\rm g}} + 
\kappa _{{\rm d}} \rho _{{\rm d}} \right) l_{{\rm BE}},
\end{equation}
where $\kappa _{{\rm g}}$ and $\kappa _{{\rm d}}$ denote the Planck 
mean opacities of gas and dust, 
respectively,\footnote{
We should note that here the definition of dust Planck mean opacity
$\kappa _{{\rm d}}$ is different from that of $\kappa _{{\rm gr}}$ 
in \citet{Omukai00} in that the former is defined as the absorption 
cross section per unit dust mass, while the latter is defined as the 
absorption cross section per unit gas and dust mass.}
$\rho _{{\rm g}}$ denotes the mass 
densities of gas ($\rho = \rho _{{\rm g}} + \rho _{{\rm d}}$) and
$l_{{\rm BE}}$ is the Bonnor-Ebert length as the typical size of the 
cloud.
Planck-mean opacities of dust $\kappa _{{\rm d}}$ are 
calculated using optical constants from references tabulated in
\citet{Nozawa08} and shown in Figure \ref{fig:Tkappa} 
for grain species considered in this study.
We then obtain the escape fraction of continuum emission as 
$\beta _{{\rm cont}}=\min \{ 1, \tau _{{\rm cont}}^{-2} \}$ \citep{Omukai00}.

In this study, we primarily consider the growth of $\Pyroxene$ grains;
Mg-silicate is one of the major grain species ejected by 
core-collapse supernovae \citep{Bianchi07,Nozawa07}.
We also investigate the growth of $\Olivine $ grains
and its effects on the thermal evolution of the gas clouds
(see Section \ref{sec:Discussion} below).
We assume that the growth of these multi-element grains is regulated 
by the least abundant gaseous species among reactants, which we 
call the key element hereafter \citep[see][]{Hirashita11}.

Using the concept of the key element, we calculate grain 
growth by following the prescription in \citet{Nozawa12}.
On the assumption that pre-existing dust grains 
are spherical with a single initial radius $r_{i,0}$, 
the evolution of grain radius, 
$r_{i}$, is described by 
\begin{equation}
\frac{dr_i}{dt} = s_i \left( \frac{4\pi}{3} a_{i,0}^3 \right)
\left( \frac{kT_{{\rm g}}}{2\pi m_j} \right)^{1/2}
n_j^{{\rm gas}} (t),
\label{GG:drdt}
\end{equation}
where $s_i$ denotes the sticking probability of the gaseous atoms incident 
onto grain surfaces, $a_{i,0}$ are the average radius of a monomer molecule 
in the dust phase {\it per key element}, and $m_j$ and $n_j^{{\rm gas}}$ are 
the mass and the number density of key element $j$, respectively.
In equation (\ref{GG:drdt}), the evaporation term can be neglected.
When the cloud fragmentation is triggered through efficient 
cooling by dust, the temperature of dust increases only up to several hundred 
kelvin, well below the sublimation temperature of $\Pyroxene $ 
($\sim$1000 K under the condition considered here). 
Therefore, dust evaporation does not affect the fragmentation condition in the
situations we investigate.

The time evolution of condensation efficiency $f_{ij}$, defined as the 
number fraction of the key element $j$ locked in grain species $i$, is
described by $f_{ij}(t) = f_{ij,0} [r_i(t)/r_{i,0}]^3$,
where $f_{ij,0}$ represents the initial condensation factor.
Then, the mass density of the grains is given by 
$\rho_{{\rm d},i}(t) = \rho X_{{\rm H}} A_j f_{ij} \mu _i$,
where $X_{{\rm H}}$ is the mass fraction of hydrogen nuclei (set 0.76 here), and
$\mu_i$ is the molecular weight of grains per key element.

We set the initial density and temperature of a gas cloud 
to be $n_{{\rm H},0}=3000 \ \percc$ and $T_{{\rm g},0}= 200 \ {\rm K}$, 
respectively.\footnote{The initial condition of the gas cloud is taken from the values 
when a supernova shell becomes gravitationally unstable in our 
one-dimensional calculations of a supernova remnant \citep{Chiaki12}.
We have confirmed that the evolutions of collapsing gas clouds at 
$n_{{\rm H}}>10^8 \ \percc $ are insensitive to these initial values.}
The initial number fractions of chemical species relative to hydrogen nuclei are
$A_{{\rm e}} = 10^{-5}$, $A_{{\rm D}}=2.87\E{-5}$, 
$A_{{\rm H}_2} = 10^{-3}$, $A_{{\rm HD}} = 10^{-6}$,
irrespective of the metallicity $Z$.
The number abundance of an element $j$ is given as 
$A_j=A_{j,\odot} (Z/Z_{\odot })$, where
the solar number abundances of heavy elements, $A_{j,\odot}$, are 
taken from \cite{Caffau11Nat}.

The average radius of an atom per key element 
in the dust phase is taken from \citep{Nozawa03}, and 
the sticking coefficient is assumed to be unity.
For the scaled solar elemental abundances
($A_{\rm Mg}/A_{\rm Si} > 1$), the key element of 
$\Pyroxene$ is Si, whose initial condensation factor is taken as
$f_{\Pyroxene,{\rm Si},0}=0.001$, 0.01, and 0.1.
As for the initial radius of pre-existing grains, we consider four 
cases of $r_{\Pyroxene, 0}=0.001$, 0.01, 0.1, and 1 $\um$.
Dust temperature is calculated from the balance equation 
between heating due to collisions with the gas
and cooling by dust thermal emission.
Also, the formation rate of hydrogen molecules on grain surfaces 
is calculated by considering impact of hydrogen atoms onto grain 
surfaces as in \citet{Schneider06}.


\section{RESULTS}
\label{sec:Results}

Figure \ref{fig:nT} shows the evolutions of 
the condensation efficiency $f_{\rm \Pyroxene,Si}(t)$, 
the radius $r_{\rm \Pyroxene}(t)$, and 
the gas temperature $T_{{\rm g}}$ 
for the initial condensation factor $f_{{\rm \Pyroxene,Si},0} = 0.1$
and the initial grain radius $r_{\rm \Pyroxene,0} = 0.01$ $\um$ 
in gas clouds with metallicities $10^{-6}$--$10^{-4} \ Z_{\odot}$.
For the models with grain growth, the condensation 
efficiency (mass density of grains) increases at 
$n_{{\rm H}} = 10^{10}$, $10^{12}$, and $10^{14} \ \percc$ for
$Z = 10^{-4}$, $10^{-5}$, and $10^{-6} \ \Zsun$, respectively.
It eventually converges to unity regardless of metallicities.
This is because the timescale characterizing grain growth eventually becomes 
smaller than the timescale characterizing the free-fall collapse 
\citep[see][]{Nozawa12}.
The grain radius also reaches to a single value 
$r_{\Pyroxene , 0}(1/f_{\Pyroxene , {\rm Si}, 0})^{1/3}$.
Since the increase in grain mass density causes enhanced dust cooling,
the models with grain growth lead to lower gas temperatures than
the models in which the condensation efficiency is constant 
(i.e., without grain growth).

The circles in Figure \ref{fig:nT} depict the phase (on each trajectory)
where the cloud is expected to fragment
\citep[see][for the detailed description of the criterion for fragmentation]{Schneider10}.
Interestingly, for $Z=10^{-5} \ \Zsun $, the trajectory reaches this phase with grain growth,
whereas, without grain growth, the trajectory does not.
We therefore conclude that grain growth can reduce the metallicity (Si abundance) 
above which low-mass fragments can form.

\subsection{Critical Metal Abundance}
\label{subsec:Crit}

Figure \ref{fig:ox} shows fragmentation properties for each model we investigate.
The double and filled circles depict the models for which the fragmentation 
conditions are met when we consider grain growth, and 
double circles depict the models for which the conditions are met without grain growth. 
From these results, we can discuss the critical condition for the formation of low-mass fragments.

First, we make the comparison of our results without 
grain growth with previous works.
As the minimal condition of fragmentation, \citet{Schneider12Crit} 
found the critical dust-to-gas mass ratio 
$\Dcrit = 4.4\E{-9}$ ( $= f_{{\rm dep}} Z$), above which the dust-induced fragmentation occurs.
In the dust model they used,
the grains have size distribution functions
with the peak at $0.01 \ \um$.
Thus, we can compare their result with our models for 
$r_{\Pyroxene,0}=0.01 \ \um$.
To apply this, we should convert $f_{\Pyroxene ,{\rm Si},0}$ into
$f_{{\rm dep}}$ as
\begin{equation}
f_{{\rm dep}} 
= \frac{ \rho _{{\rm d}, \Pyroxene} }{ \rho _{{\rm metal}} }
= \frac{ X_{{\rm H}} A_{{\rm Si}} \mu _{\Pyroxene} }{Z} f_{\Pyroxene ,{\rm Si},0},
\end{equation}
where $\rho _{{\rm metal}}$ is the mass density of metal in both gas and dust phase
and $\mu _{\Pyroxene }= 100$.
With the solar metallicity $\Zsun=0.0153$ \citep{Caffau11Sol}, we 
obtain $f_{{\rm dep}} = 0.16 f_{\Pyroxene ,{\rm Si},0}$.
In Figure \ref{fig:ox}, we plot
$Z=\Dcrit /(0.16 f_{\Pyroxene ,{\rm Si},0})$ 
by the dashed line.
Our results are consistent with the critical dust-to-gas mass ratio 
found by \citet{Schneider12Crit}.

Then, we discuss how the critical condition is changed if we consider grain growth.
The solid line in Figure \ref{fig:ox} shows the critical line above which
the fragmentation condition is met with grain growth.
Comparison of the two critical lines (solid and dashed line) shows that lower metallicity is 
generally allowed for low-mass fragment to form when we consider grain growth.
Also, the solid line in Figure \ref{fig:ox} shows that
the critical metallicity is insensitive to the initial condensation factor.
This is because even if the initial condensation factor is small 
($f _{\Pyroxene, {\rm Si}, 0} \sim 0.001$), the grains grow up to unity before the 
dust emission cooling becomes effective
unless the metallicity is too small.

Instead of the condensation efficiency (dust-to-gas mass ratio), the critical condition
is determined by the metallicity and dust size.
The estimated critical metallicity is 
$\Zcrit \simeq 10^{-5.5} \ \Zsun$ 
for the initial radius of $\Pyroxene$ grains $r_{\Pyroxene ,0} \leq 0.01 \ \um $, and 
$\Zcrit \sim 10^{-4.5} \ \Zsun$ for $f_{\Pyroxene, {\rm Si} ,0} \geq 0.1$ $\um$.
For larger initial radius, the gas density 
at which grains grow is higher
because the total surface area of dust 
particles, which determines the growth rate, is 
proportional to $r_i ^{-1}$ (for a 
fixed mass density of dust).
The right panel of Figure \ref{fig:ox} shows that 
for $r_{\Pyroxene ,0}=0.001$ and $Z=10^{-3} \ \Zsun$, the 
fragmentation conditions are not met without grain growth, since 
heating due to H$_2$ formation compensates for cooling by dust emission.

\begin{figure}[t]
\epsscale{1.0}
\plotone{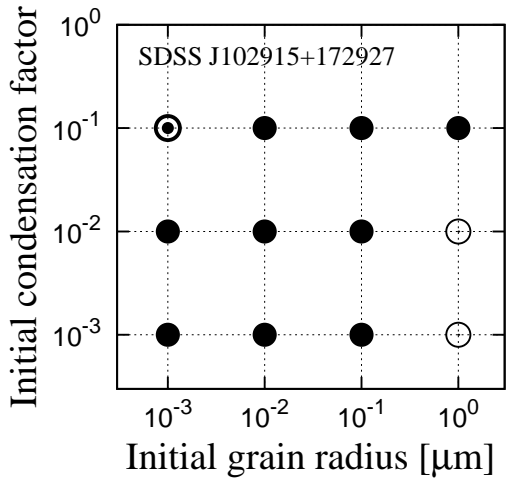}
\caption{
Fragmentation conditions of collapsing gas clouds
on $f_{\Pyroxene ,{\rm Mg},0}$-$r_{\Pyroxene ,0}$ plane
for the abundance of heavy elements of \Caf.
Symbols represents the same meaning as 
Figure \ref{fig:ox}.}
\label{fig:ox_Caf}
\end{figure}

\subsection{Application to \Caf }

We perform the 
calculations of grain growth and evolution of collapsing 
gas, adopting the abundance of heavy elements for the star \Caf
:
$A_{{\rm C}}=1.58\E{-8}$,
$A_{{\rm O}}=2.51\E{-8}$,
$A_{{\rm Mg}}=8.91\E{-10}$,
$A_{{\rm Si}}=1.78\E{-9}$, and
$A_{{\rm Fe}}=3.39\E{-10}$,
corresponding to the metallicity $4.5 \E{-5} \ \Zsun$.
Note that the abundance of carbon is an upper limit 
because no strong carbon enhancement is evident in the stellar spectrum.
Also, they estimate the oxygen abundance, 
assuming the typical excess of the  $\alpha $-element oxygen \citep{Caffau11Nat}.
Given the abundance of the gas, we aim at
constraining the initial condensation 
factor and the initial grain radius 
necessary for triggering the formation of the low-mass star.

Figure \ref{fig:ox_Caf} shows the results of
calculations with the growth of
$\Pyroxene $ grains. In the present case, the key element is Mg
because of $A_{{\rm Mg}} / A_{{\rm Si}} <$ 1.
Interestingly, for most of the cases considered here
(all cases except for $f_{\Pyroxene ,{\rm Mg},0} =$ 0.01 and 0.001 
with $r_{\Pyroxene ,0} = 1$ $\mu$m),
the fragmentation conditions are met.
Thus, we argue that the parent cloud of 
\Caf~could have been enriched with relatively small grains
($r_{\Pyroxene ,0} \la 0.1$ $\mu$m) or otherwise with relatively
high condensation factor of dust ($f_{\Pyroxene ,{\rm Mg},0} \ga 0.1$).

It is worth mentioning that the above results are approximately 
consistent with those obtained with the scaled solar composition 
(Figure \ref{fig:ox}), although the abundance pattern is different.
The results for $r_{\Pyroxene ,0} = 0.01 \ \um $ in Figure \ref{fig:ox_Caf} 
lie between those with metallicity $\log(Z/Z_\odot) = -5$ 
and $-4$ in the left panel of Figure \ref{fig:ox}, 
and the results for $f_{\Pyroxene ,{\rm Mg},0} = 0.1$ lie between 
those with metallicity $\log(Z/Z_\odot) = -5$ and $-4$ in the 
right panel of Figure \ref{fig:ox}.

\section{DISCUSSION}
\label{sec:Discussion}

In our calculations for the scaled solar abundances, 
the least abundant element, silicon, regulates the growth of $\Pyroxene $ grains.
Though we have assumed that Si atoms are directly accreted onto the grains, 
a part of or all the Si atoms might be in the form of SiO molecules in dense clouds.
We also examine a model in which all Si atoms are in the form of SiO molecules initially.
As a result, the density at which grains grow rapidly and the resulting
critical metallicity are not changed from 
the results of models in which we consider Si atoms
as the key element.

We also examine the growth of other grain 
species and the effect of each species on the thermal evolution of clouds.
We find that, for the scaled solar composition
($A_{\rm Si}/A_{\rm Fe}=1$),
$\Magnetite $ grains grow more slowly than $\Pyroxene $, 
and the gas density at which the condensation 
efficiency of $\Magnetite $ 
grains becomes 0.5 is higher than for 
$\Pyroxene$ by a factor of $\sim$5.
This is because the volume of a monomer of $\Magnetite $ grains 
($\propto a_{\Magnetite ,0}^3$) is smaller than $\Pyroxene $ and
thus the rate of grain growth is smaller (see Equation \ref{GG:drdt}).
This is also true for $\Olivine$ grains 
for which the volume per key element, Mg
atom, is smaller than for $\Pyroxene $.
However, for these grain species, the 
critical metallicity is the same as 
for $\Pyroxene $ grains
since these grain species have similar mean
opacities (Figure \ref{fig:Tkappa}) and thus similar cooling efficiencies through 
thermal emission.

On the other hand,
Si grains have mean opacity smaller than $\Pyroxene $ by orders 
of magnitude (Figure \ref{fig:Tkappa}).
Thus, even if the condensation efficiency reaches unity,
the gas cloud cannot cool down enough to meet the 
fragmentation conditions even for 
metallicities $Z \simeq 10^{-2} \ \Zsun$.
The growth of Fe grains results in larger critical abundance than $\Pyroxene $
because the species also have smaller opacity than $\Pyroxene $.
Hence, these grain species are unlikely to
affect heavily the evolution of clouds.
Indeed, Si and Fe grains could be rapidly 
oxidized into silicate and magnetite in a gas cloud with 
oxygen-rich composition.
Supernova models (N. Tominaga et al. 2013, in preparation) and observations 
in the Galactic halo \citep{Yong12} have suggested
that the abundances of oxygen 
in metal-poor stars are usually 
$A_{\rm O} / A_{X} > 10$ where $X = {\rm Mg}$, Si, and Fe.

It is considered that C grains might have also large effects
on the evolution of collapsing clouds.
Yet, it is uncertain whether C grains can grow 
sufficiently in the collapsing gas because a large 
fraction of carbon atoms are expected to be locked in 
CO molecules and are not available for the growth of C grains
\citep{Nozawa12}.
Even if CO molecules are accreted onto dust nuclei, 
the CO ice mantle would be easily 
evaporated because of its low sublimation temperature.
Nevertheless, if the initial condensation factor of
C grains is high enough, the cooling of the gas 
by carbon dust thermal emission may have an impact on the 
cloud evolution.

We have examined the effect of growth of an individual dust species 
on collapsing gas clouds separately.
In a low-metallicity gas, multiple dust species
would grow simultaneously, and thus their combined effects could 
reduce further the critical metallicity.
Dust formation calculations in primordial supernovae
\citep[e.g.,][]{Todini01,Nozawa03,Schneider06,Bianchi07}
showed that various grains species can be produced.
It is important to study how the growth of multiple species of dust 
affects the thermal evolution of collapsing gas clouds 
based on the realistic supernova dust models.

Finally, we have set the sticking probability $s_i = 1$ in the
calculations. As seen from Equation (\ref{GG:drdt}), a lower sticking
probability leads to a lower growth rate of dust. We perform another calculations,
setting $s_i = 0.1$. As a result, the fragmentation fails for $Z =
10^{-5} \  \Zsun$ and $r_{\Pyroxene, 0} = 0.01 \ \um$. However, even
for $s_i = 0.1$, grains with $r_{\Pyroxene , 0} = 0.001 \ \um$ rapidly
grow to cause the cloud fragmentation. We have considered
dust grains with a single initial size, but dust grains that
were formed and processed in supernovae may have
size distribution. Since smaller grains grow more rapidly,
the lower limit of size distribution and the initial mass
fraction of the small grains could be crucial quantities for
cloud fragmentation. 
We will investigate these effects comprehensively in our future work.

\acknowledgments

We thank K. Omukai, R. Schneider, 
N. Tominaga, and Beate Patzer for fruitful discussion.
This work is supported in part by the Grants-in-Aid for Young Scientists 
(S: 20674003, A: 22684004) 
and for General Scientists (S: 23224004) by the Japan Society for the 
Promotion of Science. 
N.Y. acknowledges financial support from World Premier
International Research Center Initiative (WPI Initiative), MEXT, Japan.


\end{document}